\documentclass{iau-JDSS}
\usepackage{graphicx}

\title[Multiple Sequences of M-dwarfs in Globular Clusters: an Infrared View.] 
{Multiple Sequences of M-dwarfs in NGC\,2808 and $\omega$ Centauri.}

\author[A.\,P.\, Milone]   
{A.\,P.\, Milone$^1$%
}

\affiliation{Instituto de Astrof\`\i sica de Canarias and Department of Astrophysics, University of La Laguna, E-38200 La Laguna, Tenerife, Canary Islands, Spain;\break email: milone@iac.es\\
}

\pubyear{2009}
\volume{Volume 15}  
\pagerange{119--126}
\date{?? and in revised form ??}
\jname{Highlights of Astronomy, Volume 14}
\editors{Ian F Corbett, ed.}
\begin{document}

\maketitle

\begin{abstract}
The infrared channel of the Wide-Field Camera 3 on the {\it Hubble Space Telescope} revealed multiple main sequences of very low-mass stars in the globular clusters NGC\,2808 and $\omega$\,Cen. In this paper I summarize the observational facts and provide a possible interpretation.
\end{abstract}
\keywords{stars: population II, globular clusters: individual (NGC\,2808, $\omega$~Centauri).}

\firstsection 
\section{Introduction}
In the context of multiple stellar populations, NGC\,2808 and $\omega$\,Cen are certainly two of the most intriguing objects. 
The CMD of NGC\, 2808 shows three main sequences (MSs), with middle and blue MS being highly helium enhanced  up to $Y \sim$0.39 with respect to the red MS  which has primordial helium (D'Antona et al.\ 2005, Piotto et al.\ 2007). 
Furthermore, spectroscopic studies have revealed significant star-to-star variations in the light-element abundances with the presence of an extreme Na-O anticorrelation (e.g.\, Norris 1981, Carretta et al.\ 2006).

The observational scenario for $\omega$\,Cen is even more complex. 
Photometry shows a multimodal MS (Bedin et al.\ 2004, Bellini et al.\ 2010), which imply extreme helium enhancement (D'Antona et al.\ 2004, Norris et al.\ 2004).
At odds with most globulars, which have homogeneous iron abundance, 
 its stars span a wide interval of metallicity and define a multimodal distribution in [Fe/H], and $s$-elements. Large star-to-star light elements variations, Na-O and C-N anticorrelations are present within each metallicity interval (e.g.\ Marino et al.\ 2011). 

Photometry of globular-cluster sequences extended to date over a limited spectral region, from the ultra-violet (UV, $\lambda \sim$ 2000\AA) to the near-infrared (NIR, $\lambda \sim$ 8000\AA).  As such, multiple sequences are rarely detected along the lower part of the MS, because observational limits make it hard to get high-accuracy photometry of very faint and red stars in optical and UV colors. 
In the following we use {\it HST} to extend the study to the NIR passbands and investigate multiple sequences in NGC\,2808 and $\omega$\,Cen over a wide interval of stellar masses, from the turn off down to very low-mass MS stars ($\mathcal{M} \sim$0.2$\mathcal{M}_{\odot}$).     

\section{Multiple populations of very low-mass stars}
Figure~1a shows the NIR CMD for NGC\,2808 from Milone et al.\ (2012a, left panel) and $\omega$\,Cen (right panel). 
The upper MS of NGC\,2808 is consistent with three stellar populations with different helium and light-element abundance, in agreement with previous observations based on visual photometry. The three MSs merge together at the luminosity of the MS bend while at fainter magnitudes, at least two MSs can be identified. A bluer, more populated $MS_{\rm I}$, which includes $\sim65$\% of MS stars, and a $MS_{\rm II}$ with $\sim35$\% of stars. The fractions of stars along $MS_{\rm I}$, and $MS_{\rm II}$ are very similar to the fraction of red-MS stars ($\sim$62\%) and the total fraction of middle-MS and blu-MS stars ($\sim$24+14=38\%, Milone et al.\ 2012b).

The observed CMD of NGC\,2808 has been compared with appropriate evolutionary models for very low-mass stars and synthetic spectra that account for the chemical composition of the three stellar populations of this clusters (see Milone et al.\ 2012a for details).
It comes out that $MS_{\rm I}$  is associated with the first stellar generation,
which has primordial He, and O-C-rich/N-poor stars, and that  $MS_{\rm II}$,  corresponds to a second-generation stellar population that is enriched in He and N and depleted in C and O.  
The $MS_{\rm I}$ of Fig.~1a is the faint counterpart of the red MS identified 
by Piotto et al.\ (2007), whereas the $MS_{\rm II}$ corresponds to the lower-mass counterpart of the middle MS and blue MS of Piotto et al.\ paper.  

The brightest part of the CMD of $\omega$\,Cen in Fig.~1b shows that the blue and the red MS, already detected in previous papers, merge together at the magnitude of the MS bend. At fainter luminosities, it appears a broad MS with a blue, more populated component ($MS_{\rm I}$), and a red tail ($MS_{\rm II}$).
The similarities with NGC\,2808, make it very tempting to associate the $MS_{\rm I}$ and the $MS_{\rm II}$ to the red and the blue MS, respectively. A comparison of the CMD with stellar models and synthetic spectra that account for the complex chemical composition of $\omega$\,Cen stellar populations is mandatory to clarify this issue.   
\begin{figure}
\begin{center}
\includegraphics[height=2.55in,angle=0]{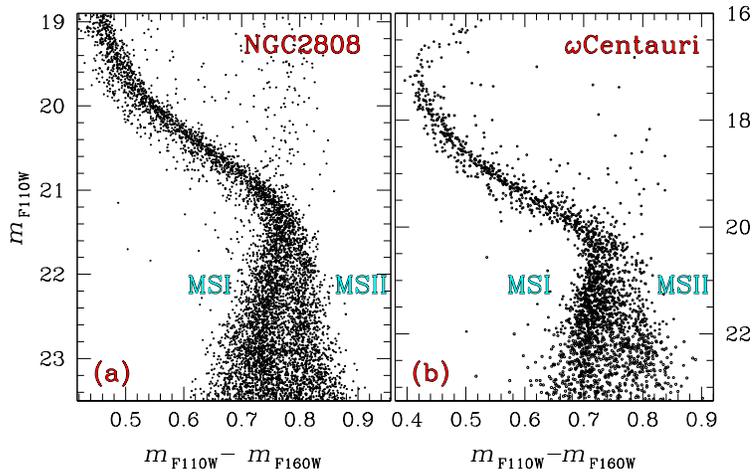}
\caption{NIR CMD corrected for differential reddening for NGC\,2808 and $\omega$ Cen.}\label{fig1}
\end{center}
\end{figure}

These results provide the first detection of multiple populations with different helium and light-element abundances among very low-mass stars. The fact that the signatures of abundance anticorrelation are also observed among fully-convective M-dwarfs definitely demonstrates that they have primordial origin and hence correspond to different stellar generations. 

\begin{acknowledgements}
I warmly thank A.\ F.\ Marino, S.\ Cassisi, G.\ Piotto, L.\ R.\ Bedin, J.\ Anderson, F.\ Allard, A.\ Aparicio, A.\ Bellini, R.\ Buonanno, M.\ Monelli, A.\ Pietrinferni for their collaboration to this work.
Support for this work has been provided by the IAC (grant 310394), 
and the Education and Science Ministry of Spain (grants AYA2007-3E3506, and AYA2010-16717).
\end{acknowledgements}

\end{document}